**Joseph F. Ryan***

Sars International Centre for Marine Molecular Biology, University of Bergen, Bergen, Norway

Current Address:

Whitney Laboratory for Marine Bioscience, University of Florida, St. Augustine, FL, USA

E-Mail: joseph.ryan@whitney.ufl.edu

**Abstract**

Assessing the correctness of genome assemblies is an important step in any genome project. Several methods exist, but most are computationally intensive and, in some cases, inappropriate. Here I present baa.pl, a fast and easy-to-use program that uses transcript data to evaluate genomic assemblies. Through simulations using human chromosome 22, I show that baa.pl excels at detecting levels of missing sequence and contiguity. The program is freely available at: https://github.com/josephryan/baa.pl

**Introduction**

Genome assembly is becoming commonplace in modern molecular biology labs. Generating multiple assemblies with different sets of parameters in different programs, then comparing these assemblies is a key step in genome assembly. However, comparing genomic assemblies is challenging since gold standards are naturally missing. The most common method of assembly comparison uses measures of contiguity as the main metric (*e.g.*, N50), though relying on levels of contiguity alone may be problematic [1]. For example, the level of contiguity of an assembly where all reads were randomly aligned end-to-end into a single sequence would be extremely high, but the assembly itself would be nonsensical.

Several techniques have been developed to assess genome assemblies. These methods either depend on the mapping of reads used to produce the assembly [2,3,4,5,6,7], or the identification and counting of highly conserved sequences [8]. These techniques are powerful—and in some ways superior to the method described here —but are computationally expensive and therefore time-consuming. In addition, because read-mapping techniques assume an even distribution of reads over the genome, they are inappropriate in cases where this assumption is violated. This is true, for example, in cases in which data was generated using whole-genome amplification, due to the introduction of PCR amplification biases in coverage.

RNA transcripts intrinsically contain information about the structure of genomes; therefore, they are an excellent source of independent information for comparing and validating genome assemblies. To facilitate this utility, I have developed the program baa.pl.

**Results**

Baa.pl requires that a set of RNA transcripts are first aligned to the genome using the freely available program BLAT [9]. The output of this program, as well as the FASTA file of the RNA transcripts used as input to BLAT, are the only required inputs to baa.pl. The output consists of four measures, explained in Table 1.

The ratio of transcripts with a BLAT entry (Table 1a) is determined by dividing the number of transcripts with a BLAT alignment by the total number of transcripts. The algorithm groups each set of sub-alignments in the BLAT file by query transcript. A score for each BLAT sub-alignment is determined using the algorithm implemented by James Kent (the author of BLAT) in the UCSC Genome Browser [10]. Starting with the highest scoring sub-alignment, each position in the query transcript that aligns is marked as "covered" using an array the size of the query sequence. Each successive sub-alignment (sorted by decreasing score) is then considered. If a sub-alignment contributes more than X positions (X by default = 5) to the coverage of the query, then the sub-alignment is considered to have contributed to the coverage. If the target sequence of the contributing sub-alignment has not previously contributed to the "coverage," then the number of target sequences (*e.g.*, contigs) contributing to the alignment is incremented. These calculations are used to determine the number of transcripts mapping to a single target sequence (Table 1b) and the average number of target sequences per mapped transcript (Table 1c). The array of covered positions is then analyzed and if more than Y consecutive positions (Y by default = 5) are not covered, these positions count against the total coverage of the query. This calculation is used to determine total percent coverage of all transcript nucleotides (Table 1d).

Baa.pl is fast compared to other assembly assessment methods. The time of the analysis is determined by how long it takes to align a set of transcripts to a genome with BLAT. Aligning transcripts to an assembly is decidedly faster than aligning all sequencing reads to an alignment or predicting highly conserved genes. The running of baa.pl itself on a BLAT output file usually takes less than 10 seconds with a reasonably sized data set and a modern computer.

To illustrate the effectiveness of baa.pl, I produced several semi-random permuted versions of human chromosome 22, then aligned mRNAs of 42 genes located on the same chromosome. Permutations to chromosome 22 included breaking it into fragments at random positions and deleting random chunks of sequence. Position 0 on the X axis of both Figure 1 and Figure 2 shows that in a perfect case scenario (*i.e.*, every gene aligns almost perfectly to the target sequence or sequences) the metrics reflect the ideal alignments of the transcripts to the genome.

To test the sensitivity of baa.pl to contiguity, I generated versions of chromosome 22 that had been divided at random positions into fragmented sequences at 1 kb frequencies between 1,000 and 10,000. The level of fragmentation is inversely analogous to the level of contiguity in a genome assembly. These results (Figure 1) show that two metrics in the form of the "number of transcripts mapping to a single sequence" and the "average number of sequences per mapped mRNA" diverge from 0 in a manner proportional to the number of breaks introduced. These results show that these two metrics are appropriate for assessing levels of contiguity. The other metrics in these breakage analyses remain mostly unchanged.

To test the sensitivity of baa.pl to missing sequence data, I introduced random sequence deletions of 5,000 nucleotides in random positions of chromosome 22 at 1 kb frequencies between 1,000 and 10,000. The number of introduced deletions is analogous to the amount of missing data in a genome assembly. In the case of deletions (Table 2c), the metrics "total percent coverage of all mRNA nucleotides" and "ratio of mRNAs with a BLAT entry" tend to decrease as more sequences are deleted. These results show that these two metrics are appropriate for assessing levels of completeness. The other metrics in these breakage analyses remain mostly unchanged.

**Discussion**

Genome sequencing has become a routine task in biology. Generating high-quality genome assemblies, however, has remained difficult. One of the major difficulties of this process is that assembly algorithms and parameter settings for these algorithms perform differently depending on the nature of the genome and the sequencing [1,11]. For this reason, it is important to run multiple assemblers with multiple sets of parameters and to integrate robust assessment methods into any genome-sequencing pipeline.

Existing tools offer metrics that can be used to compare assemblies. Methods that use alignments of sequencing reads used in constructing the assemblies are powerful, but can be misleading if the underlying assumption of uniform coverage of sequencing reads is violated. This assumption is most certainly violated in cases that use DNA amplification—a technique critical for sequencing the genomes of microorganisms. Methods that predict highly conserved genes in an assembly and compare the counts of these genes between assemblies avoid these issues, and are invaluable. However, these methods consider only a few hundred genes and therefore offer limited resolution.

The use of baa.pl to evaluate alignments of RNA transcripts offers several advantages over these methods. RNA transcripts are independently generated and, as such, are not affected by any artifacts introduced during the genomic sequencing used to generate the alignment. In addition, most eukaryotes contain tens of thousands of genes and current RNA sequencing methods recover the vast majority of these transcripts. These numbers offer a much higher resolution than a small set of highly conserved genes does. The metrics from baa.pl offer the added advantages of providing a rough assessment of the extent to which a particular genome assembly encapsulates the entire transcriptome of a species, as well as a measure of how often transcripts occur on a single sequence (an important measure in determining the extent to which gene predictions will be effective). Lastly, an analysis involving baa.pl is very straightforward and considerably faster than other assessment algorithms.

Baa.pl uses evidence from alignments of RNA transcript data to a genome assembly to generate metrics that can then be compared across multiple assemblies, each generated using different programs and/or different sets of parameters. The metrics are intuitive and offer clues as to the nature of differences between assemblies (e.g., contiguity vs. completeness). The program has no requirements outside of BLAT and Perl, and includes a straightforward installation. The program's speed, ease-of-use, and intuitive nature make it an ideal tool for genome assembly assessment.

**Materials and Methods**

I downloaded chromosome 22 and 42 mRNA products of this chromosome from NCBI on August 14, 2013. I developed a Perl script to introduce random permutations to the genomic sequence of chromosome 22 at various frequencies. I aligned the 42 transcripts to chromosome 22 and to its random permutations using default parameters in BLAT (version 35x1). I then ran baa.pl (version 0.20) with default parameters using the output of each BLAT run. All scripts and random datasets used in these analyses are available as supplemental material.

**Acknowledgements**

I thank Andreas Hejnol for providing genomic data from intriguing marine invertebrates, which motivated the development of this program; my former colleagues at the Sars Centre for their interest, insights, and encouragement; Itai Yanai, Sara Sumic, and Bernard Koch for providing feedback on previous versions of this manuscript; and Amy Datsko for copy editing. The views in this paper do not necessarily reflect the views of those acknowledged.**References**

**Tables**

**Table 1. Metrics produced by baa.pl**

| Metric | What is being assessed |
|---|---|
| **(a) Ratio of transcripts with a BLAT entry** | The number of transcripts detected that produced a BLAT alignment with the assembly divided by the total number of transcripts. **completeness** |
| **(b) Number of transcripts mapping to a single sequence** | The count of transcripts that align to a single genomic sequence (contig, scaffold or chromosome). Baa.pl will consider a genomic sequence as contributing to an alignment only if a certain number of nucleotides (specified by 'min_to_count_as_coverage' option, default=5) in an alignment block align to a region of the transcript that was not aligned in any of alignment blocks with lower E-Values. **contiguity** |
| **(c) Average number of target sequences per mapped transcript** | A count of the number of transcripts that only align to a single genomic sequence (contig, scaffold or chromosome). The length and coverage of the alignment is not considered. **contiguity** |
| **(d) Total percent coverage of all transcript nucleotides** | The nucleotides of a particular transcript that are included in one or more alignments to the assembly are considered covered. The percentage of all mRNA nucleotides is the sum of all covered nucleotides divided by the sum of all transcript nucleotides. Small gaps between alignment blocks do not contribute to the percentage of missing coverage unless they are larger than the value specified by 'max_gap_to_consider_missing' (default=5). **completeness** |

**Figures**

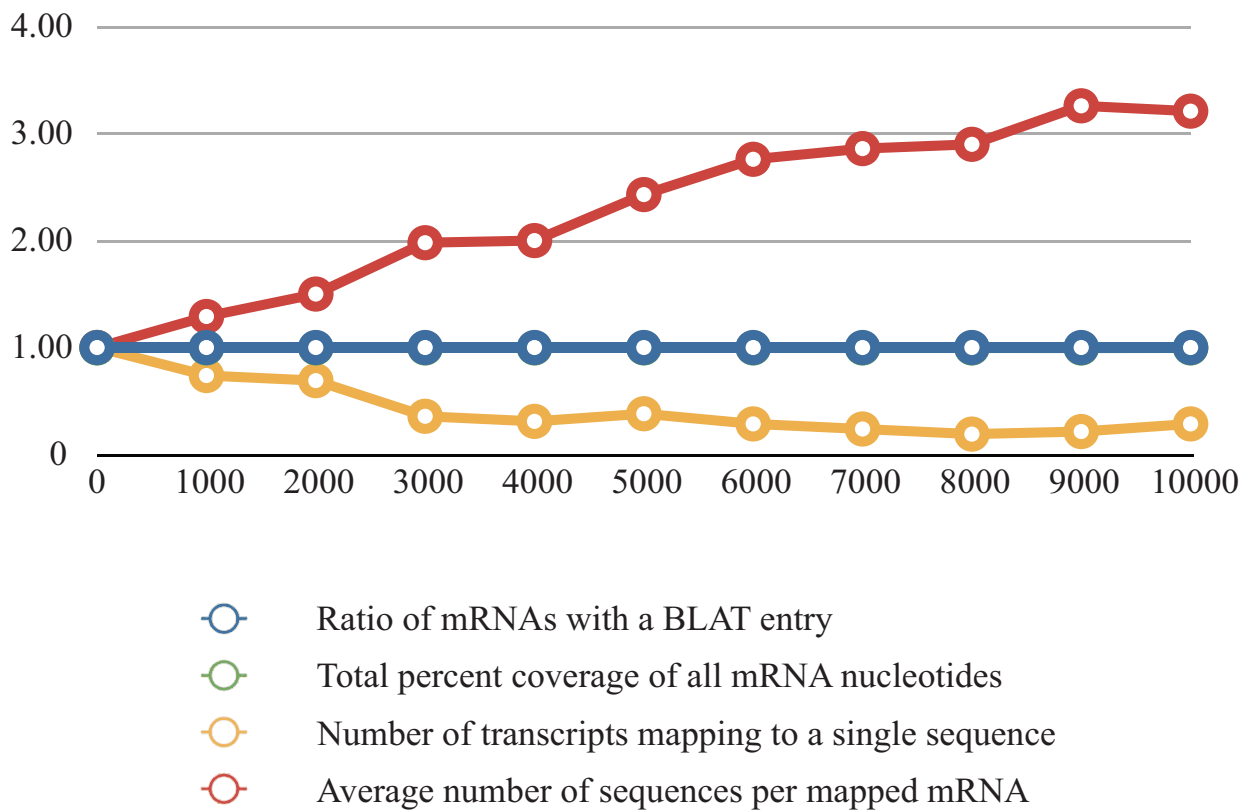

○ Ratio of mRNAs with a BLAT entry
○ Total percent coverage of all mRNA nucleotides
○ Number of transcripts mapping to a single sequence
○ Average number of sequences per mapped mRNA

**Figure 1. Effect of randomly introduced breakages of chromosome 22 on baa.pl metrics**
The average number of sequences per mapped mRNA increases as additional breakages are introduced. A related measure, the number of transcripts mapping to a single sequence decreases with the introduction of breakages. Ratio of mRNAs with a BLAT entry (blue) and total percent coverage of all mRNA nucleotides (green) are not affected.

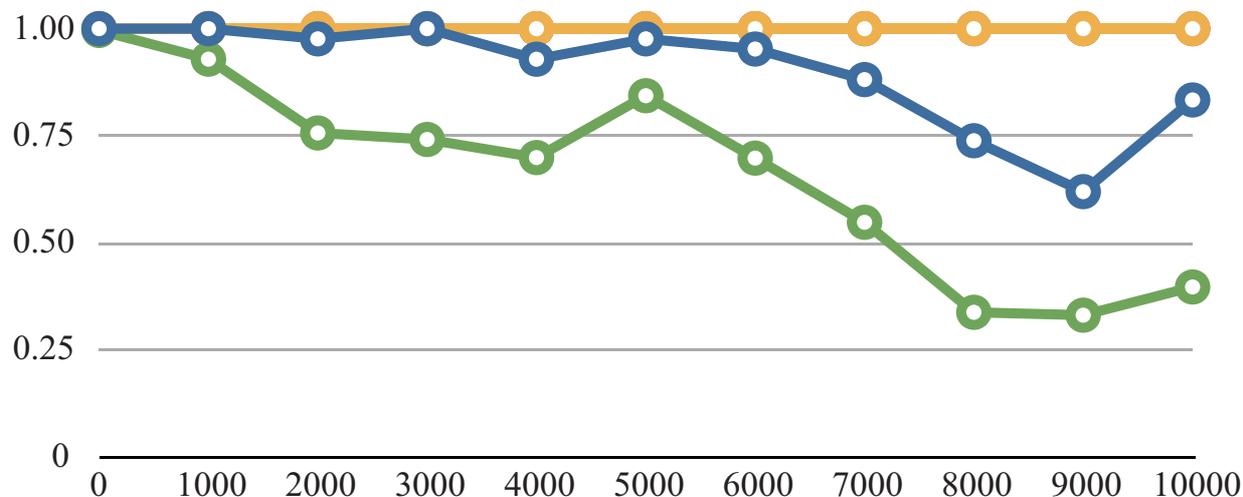

**Figure 2. Effect of randomly introduced 5KB deletions to chromosome 22 on baa.pl metrics**
The ratio of mRNAs with a BLAT entry and the total percent coverage of all mRNA nucleotides both decrease as more random deletions are introduced to chromosome 22. The number of transcripts mapping to a single sequence